\documentclass[final]{siamltex}
\usepackage{graphicx}
\usepackage{subfigure}
\usepackage{color}
\usepackage{amsmath,amssymb,amsbsy}
\usepackage{psfig}
\usepackage{epsfig}
\usepackage{algorithmic,algorithm}
\usepackage{setspace}
\def\eqn#1{\begin{equation} \label{#1}}
\def\enn{\end{equation}}
\def\eq{\[}
\def\en{\]}

\def\e#1{eq (\ref{#1})}

\def\s#1{Section \ref{#1}}
\def\f#1{Figure \ref{#1}}

\begin{document}
\title{Parameterization of non-linear manifolds}
\author{C.~W.~Gear\\
Department of Chemical and Biological Engineering\\
Princeton University, Princeton, NJ\\
e-mail:wgear@princeton.edu
}
\date{}

\maketitle


\begin{abstract}
In this report we consider the parameterization of low-dimensional manifolds that are specified (approximately) by a set of points very close to the manifold in the original high-dimensional space.  Our objective is to obtain a parameterization that is (1-1) and non singular (in the sense that the Jacobian of the map between the manifold and the parameter space is bounded and non singular).
\end{abstract}

{\bf Keywords}  Diffusion Maps, PCA, Distance Matrices

\section{Introduction}\label{introduction}

When it is known that the high-dimensional solutions of a computational problem lie on a lower-dimensional manifold we may wish to operate only on that manifold.  By repeatedly solving the problem numerically, we can compute points that are approximately on the manifold and we will assume that we are able to compute a set of such points over the manifold so that they are ``reasonably distributed" (i.e., the average density of the points does not vary too greatly from region to region.  We also assume that we can approximate any point on the manifold sufficiently accurately by interpolation from these pre-computed points and we would like to find a convenient parameterization of the manifold that can be used for the interpolation.

For example, if the solutions of a differential equation in the high-dimensional space rapidly approach a low-dimensional manifold it may be desirable to integrate a reduced system on the manifold to avoid stiffness.  In this case, it might be even more convenient to integrate in the parameter space. Note that in this example it is important that the parameterization be non singular: otherwise the differential equations in the parameter space would have a singularity.

We suppose that we are given a set of $N$ points ${\rm X}_i$, $i = 1, \cdots, N$ in $s$-dimensional Euclidean space, S, with coordinates ${\rm x}_i = \{x_{q,i}\}$, $q = 1, \cdots, s$ that are on (or close to, due to noise or computational error) a low-dimensional manifold, M.  Our ultimate objective is to find, computationally, a parameterization of any manifold that can be simply parameterized.  For now we rule out manifolds such as tori that can only be parameterized by using periodicity in some of the parameters.  A desiderata is to be able to parameterize any manifold that can be formed by ``twisting" (without any stretching) a linear manifold in the high dimensional space (such as the well-known ``Swiss roll" example in \f{F1}(b)) so that the parameterization is a good approximation to a Cartesian coordinate system for the original linear manifold.  (In that figure the X-axis is foreshortened and has a range of about 2.5 times that of the other two axes so as to show the structure.  Consequently it appears to be rolled from a fairly narrow rectangle but it is actually rolled from the square shown in \f{F1}(a).  The randomness of these points was constrained to get a ``reasonable" distribution - the 1 by 1 square was divided into 40 by 40 equal smaller squares, a point was placed in the middle of each and then subject to a random uniform perturbation in each coordinate in the range [-0.4, 0.4]/40, thus ensuring that no points were closer than 0.2/40.)

\begin{figure}[h]
  \parbox{\textwidth}{
   \includegraphics[width=.45\textwidth]{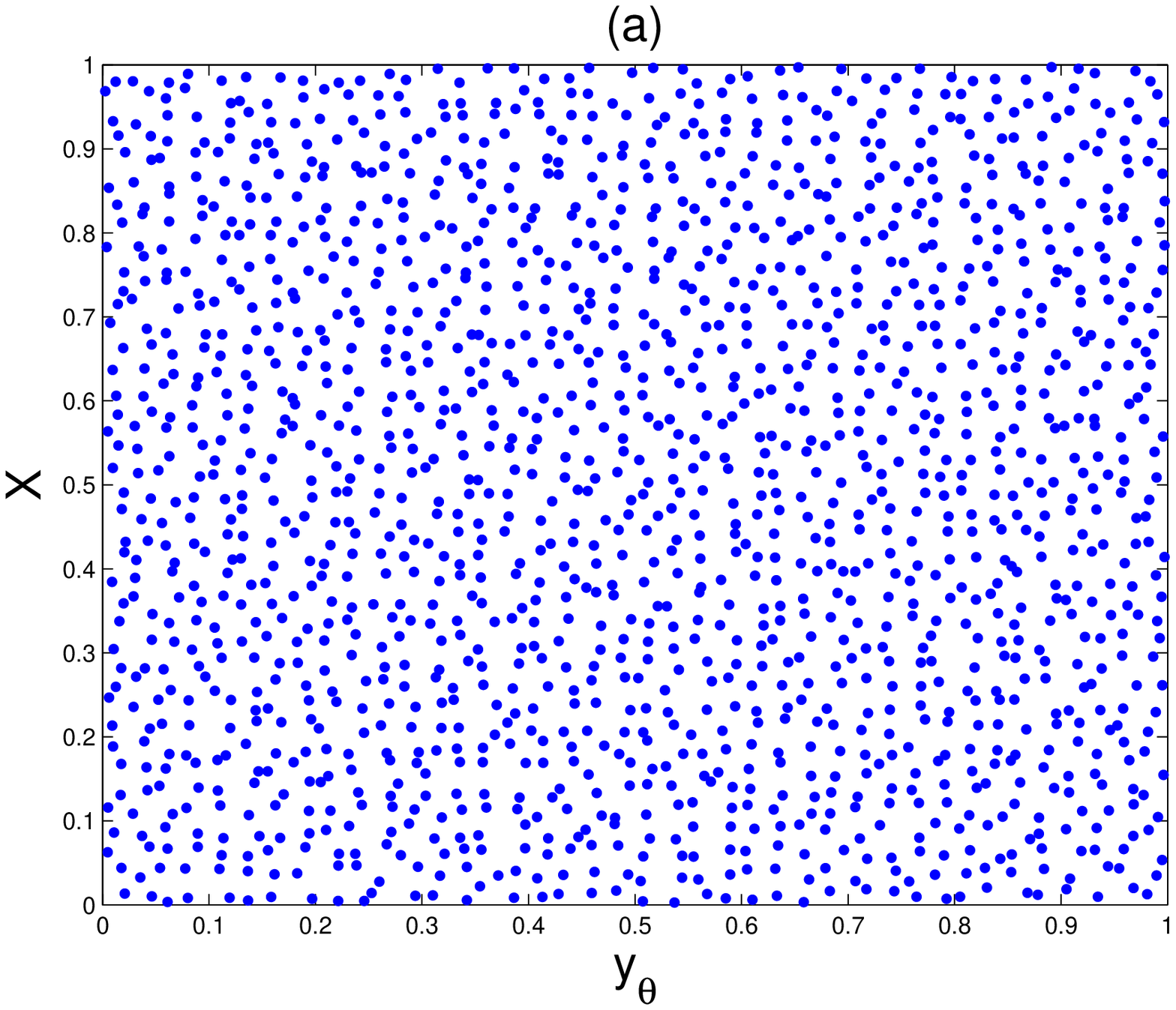}
   \includegraphics[width=.45\textwidth]{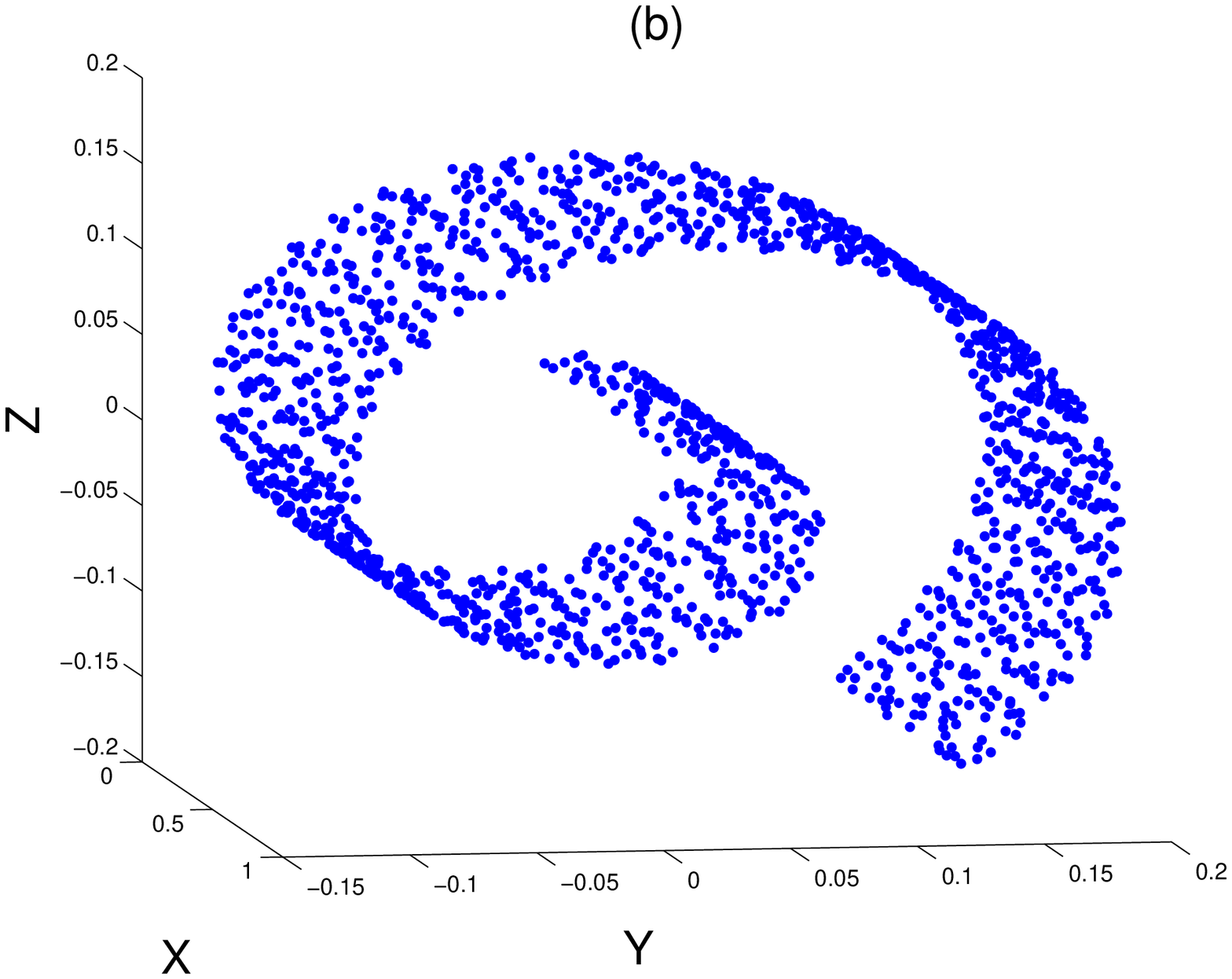}}
  \caption{(a) 1,600 points ``randomly" placed (see text) on a 1 by 1 square. (b) ``Swiss Roll" formed by rolling the square into a spiral. The $y_\theta$ coordinate in (a) correspons to distance around the spiral in (b).}
\label{F1}
\end{figure}

The learning and statistical communities have long been interested in manifold learning and representation so there is a large body of literature.  In those cases the typical task is to start with the data set $\{{\rm X}_i\}$ and produce a ``representation" in a lower-dimensional space such that the differences in the distances of pairs of points in the two spaces are suitably small.  This was achieved by minimizing a {\em stress} function in Sammon's early paper\cite{Samm69} and numerous modifications have been proposed, such as \cite{SuCrFy11,SuFyCr12}.  If the low-dimensional manifold is linear, the Principal Component Analysis (PCA) is the standard way to identify and represent it - the principal components provide an orthogonal basis for the smallest linear space containing the manifold.  Many methods for non-linear manifolds start with some modification of PCA.  For example, \cite{ZhZh05} effectively uses PCA locally and then stitches the tangent spaces together.  Local Linear Embedding (LLE) \cite{RoSa00} looks for a local linear approximation which is related to PCA.

Another class of methods is based on the structure of the graph generated by focussing attention primarily on nearby neighbors.  Diffusion Maps \cite{CoLa06} are one type of this method.  We have looked at the use of Diffusion Maps for the parameterization of slow manifolds \cite{SGSK12}.  Diffusion Maps use a parameter, $\epsilon$, that effectively specifies the distance of what is considered a nearby neighbor. When a small $\epsilon$ is used, the parameterization is nearly singular at the boundary.  A much larger $\epsilon$ sometimes overcomes this difficulty, but can lead to other difficulties.

In this report we show that that the limit of a particular case of diffusion maps for large $\epsilon$ is PCA if we use the left eigenvectors rather than the right eigenvectors normally used in diffusion maps.  However, because, in diffusion maps, we work with distances in the graph structure rather than with the coordinates $\{{\rm x}_i\}$ that are the input to PCA, we can consider modifications to the distances of pairs of points that are not nearby neighbors to improve the parameterization.

In \s{Sdemo} we discuss the problems of diffusion maps for the data set in \f{F1}(b) - for small $\epsilon$ the map has a nearly singular Jacobians, but breaks down for large $\epsilon$ because points distant in the manifold are relatively close in the high-dimensional space.  Then, in \s{S2}, we consider the large $\epsilon$ limit for diffusion maps on linear manifolds and prove that the left eigenvectors provide the PCA decomposition and show that this can be obtained by working directly with the distance matrix.
In \s{S3} we consider ways to modify the distance matrix to improve the parameterization.

\section{Diffusion Map Issues}\label{Sdemo}

Define the (symmetric) distance matrix {\bf H} to have the entries $${\bf H}_{ij} = d(X_i, X_j)$$ where $d(.,.)$ is the Euclidean distance between the points, that is
\eq
{\bf H}_{ij} = \sqrt{\sum_{q=1}^s(x_{q,i}-x_{q,j})^2}.
\en
The diffusion map method \cite{CoLa06} now forms the matrix ${\bf W} = \{{\bf W}_{ij}\} = \{w({\bf H}_{ij})\}$ where $w$ is a suitable non-negative function such that $w(0) = 1$ and $w(h)$ is monotonic non-increasing for $h > 0$ with $w(\infty) = 0$.  We will use
\eq
w(h) = \exp(-\left ( \frac{h}{\epsilon}\right )^2).
\en
The diffusion map method continues by dividing each row of ${\bf W}$ by its corresponding row sum to get a Markov matrix ${\bf K}$ whose largest eigenvalue is 1 with a corresponding eigenvector {\bf e} = $[1, 1, \cdots ,1]^T$.  Some of its next few eigenvectors serve to identify the low-dimensional manifold and the $i$-th components of these eigenvectors provide a parameterization for the points, ${\rm X}_i$, on the manifold.  It is well known that for small $\epsilon$ the parameterization is poor near the boundaries - see, for example, \cite{SGSK12} - but that a large $\epsilon$ can give a good parameterization under some circumstances.

In \f{F2} we show the results of applying a diffusion map to the data in \f{F1} (b).  Here, $\epsilon$ was chosen as $\sqrt{2}\beta$ where $\beta$ was the smallest value such that if all edges in the graph corresponding to the distance matrix ${\bf H}$ longer than $\beta$ were removed, the graph would still be connected. (For this particular data set, $\beta$ is 0.0322.)  \f{F2} (a) and (b) plot the components of the second and third eigenvectors, $\xi_2$ and $\xi_3$, against the $X$ and $y_\theta$ coordinates of the points in \f{F1}(b).  ($X$ and $y_\theta$ are the coordinates of the points in \f{F1}(a).  After the square is ``rolled" $y_\theta$ is the arc length around the spiral from its inner stating point while $X$ is the coordinate along the length of the cylindrical spiral.) We see that $\xi_2$ and $\xi_3$ come very close to providing direct parameterizations of $X$ and $y_\theta$, respectively, and this is shown more clearly in \f{F2} (c) and (d) (which are projections of the previous two figures onto axial panes).  However, we see that the parameterization is very poor near the boundary where $\partial \xi_2/\partial X$ and $\partial \xi_3/\partial y_\theta$ are nearly zero (because the curve is close to a cosine half cycle).  For example, if one were trying to integrate a system on its slow manifold using the parameterization as the dependent variables the sensitivity of the derivative to the parameter values would be very large near the boundary.

\begin{figure}[h]
  \parbox{\textwidth}{
   \includegraphics[width=.45\textwidth]{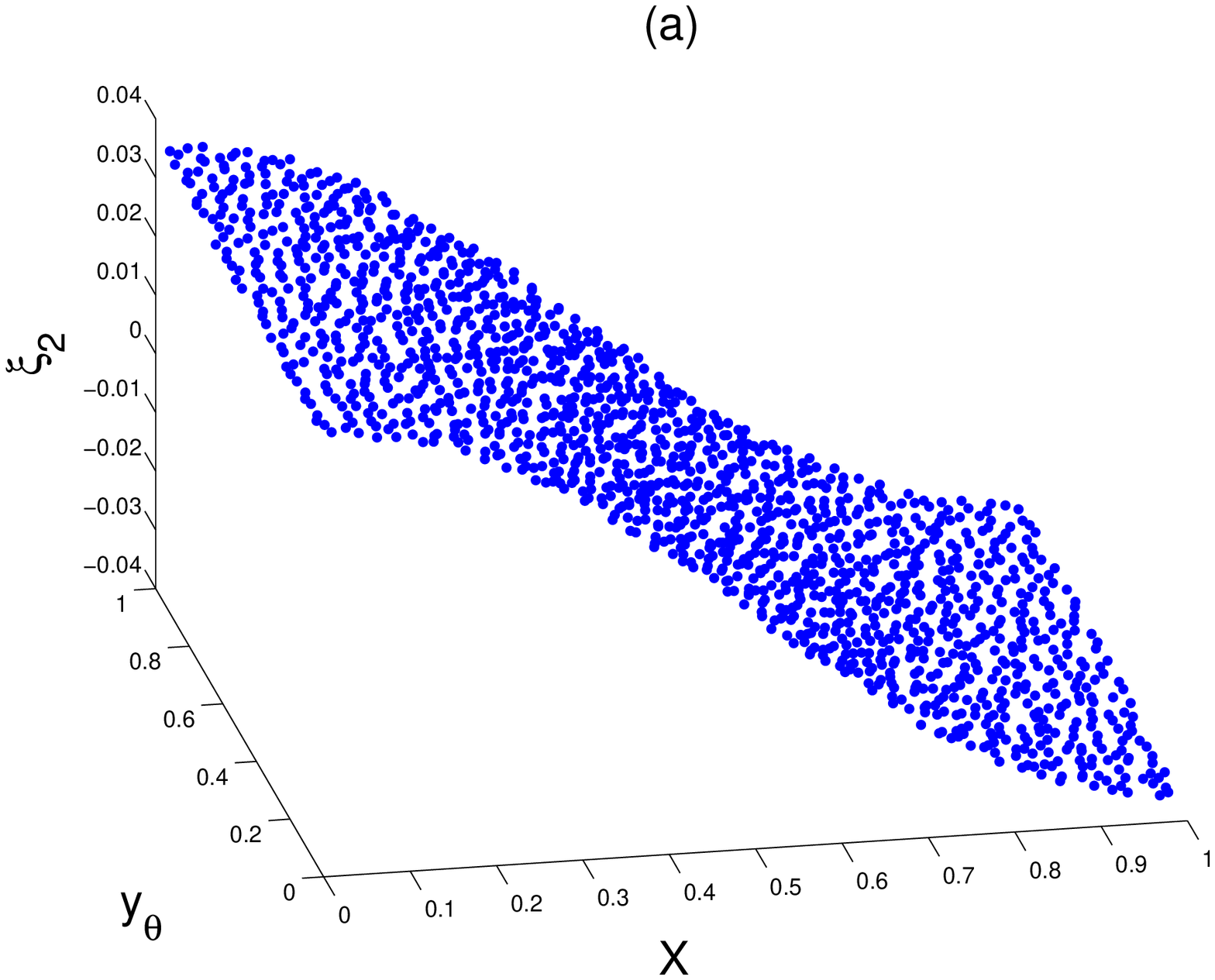}
   \includegraphics[width=.45\textwidth]{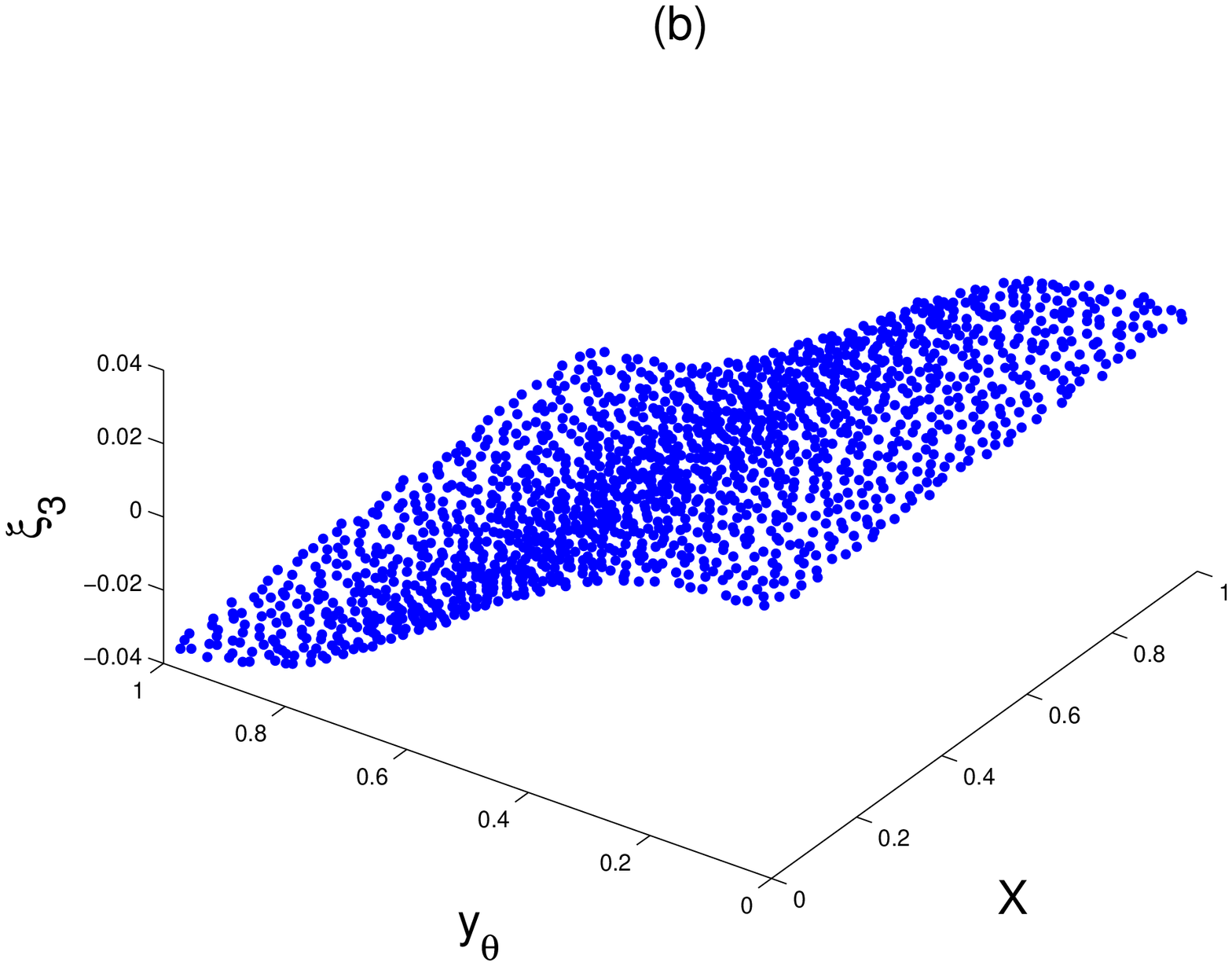}}
  \parbox{\textwidth}{
   \includegraphics[width=.45\textwidth]{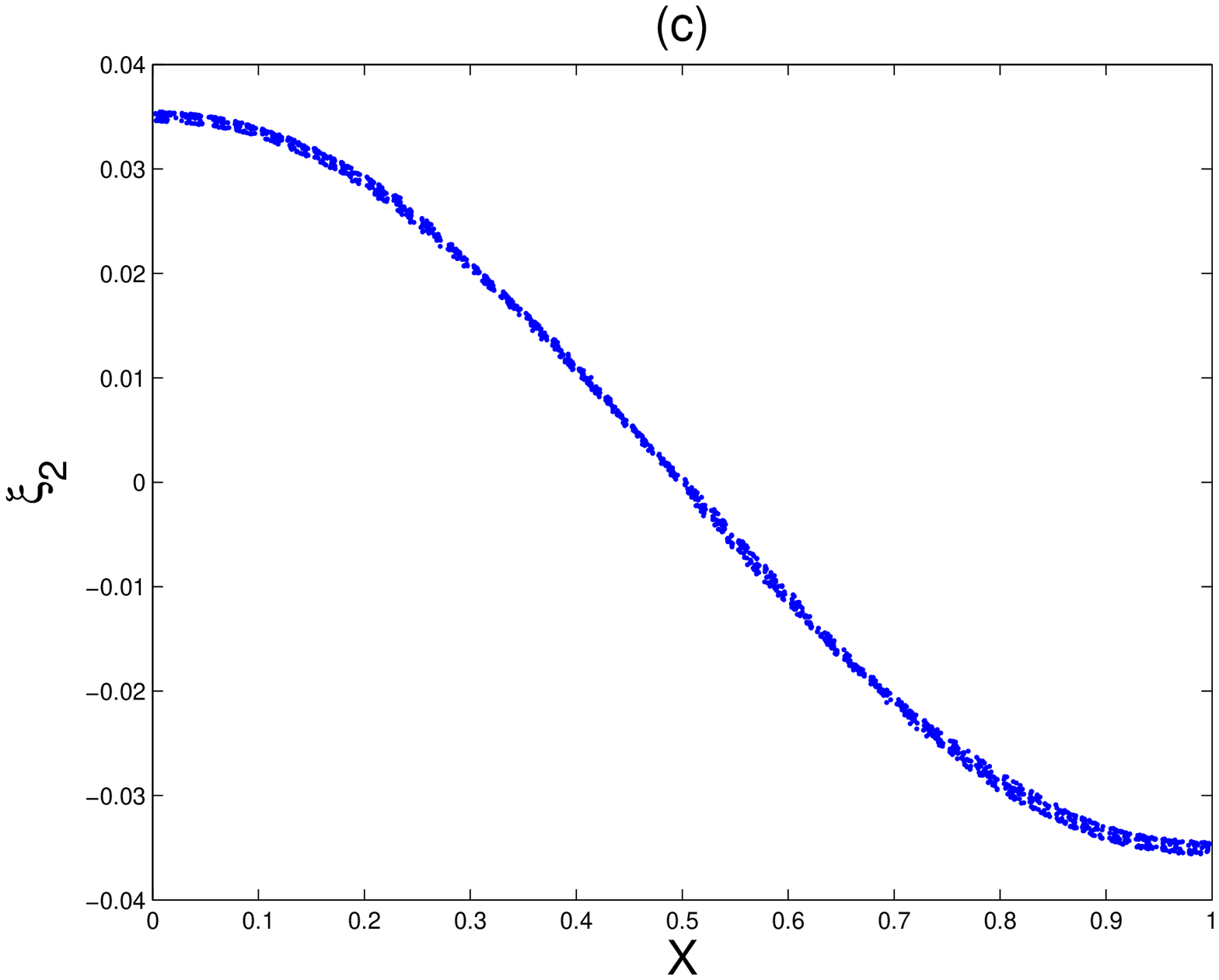}
   \includegraphics[width=.45\textwidth]{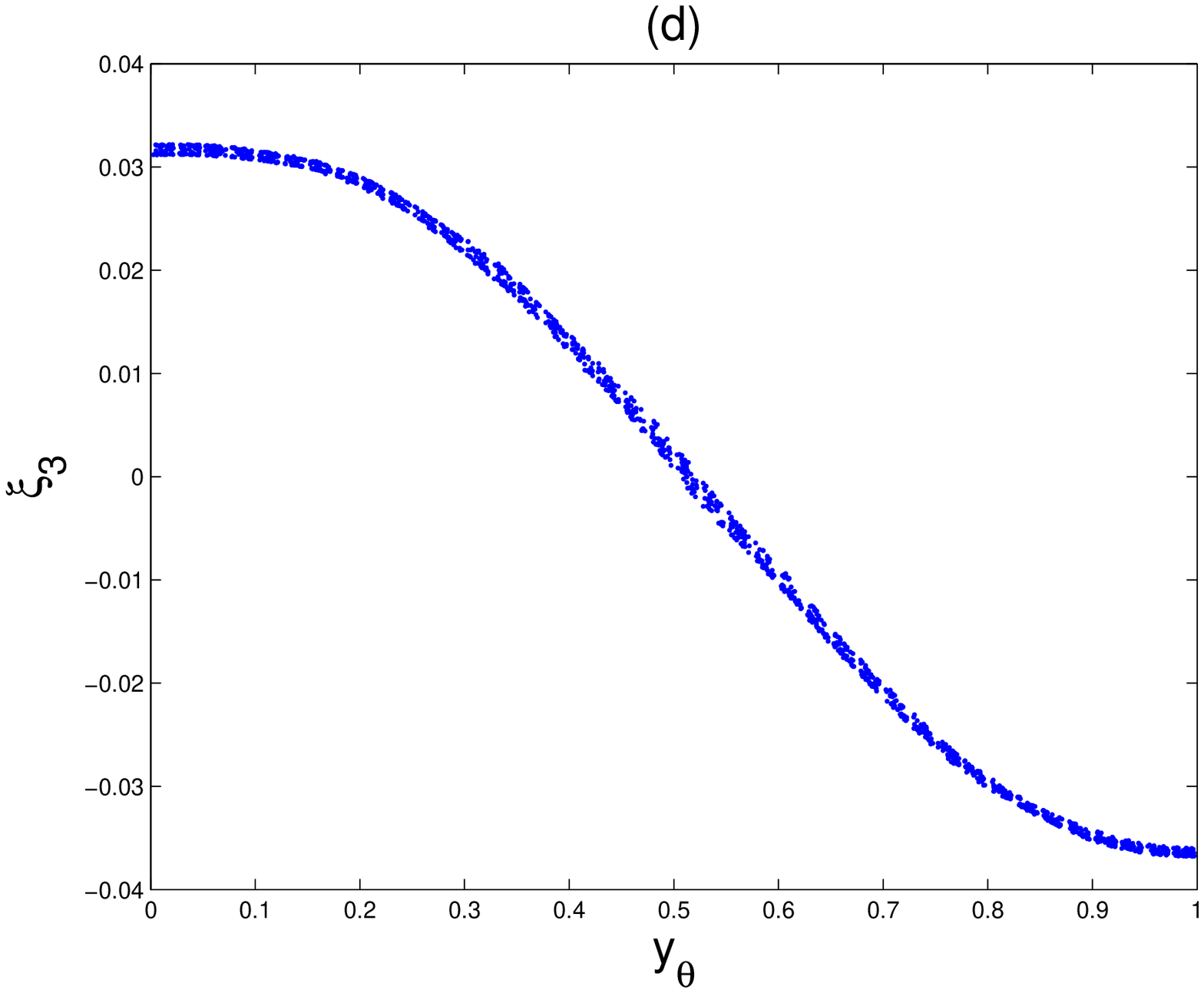}}
   \caption{Diffusion map applied to data in \f{F1}(b) with $\epsilon = 0.0455$.  (a) Second eigenvector $\xi_2$. (b) Third eigenvector $\xi_3$. (c)  $\xi_2$ versus $X$. (d)  $\xi_3$ versus $y_\theta$.}
\label{F2}
\end{figure}

In some cases, this can be overcome by using a large $\epsilon$.  In \f{F3} (a) and (b) we show the equivalent of \f{F2} (c) and (d) for a diffusion map of the same data with $\epsilon = 1$ - the distance along the cylinder.  Note that $\xi_2$ provides a very good parameterization of the $X$ direction - it is almost linear, but that $\xi_3$ is no longer a (1-1) map.  This occurs because  ${\bf W}_{ij}$ for points on adjacent turns of the spiral is about 0.985 so the points are seen as close.  (The spiral started with radius  0.03 and increased radially by 0.12 on each full turn.)

\begin{figure}[h]
  \parbox{\textwidth}{
   \includegraphics[width=.45\textwidth]{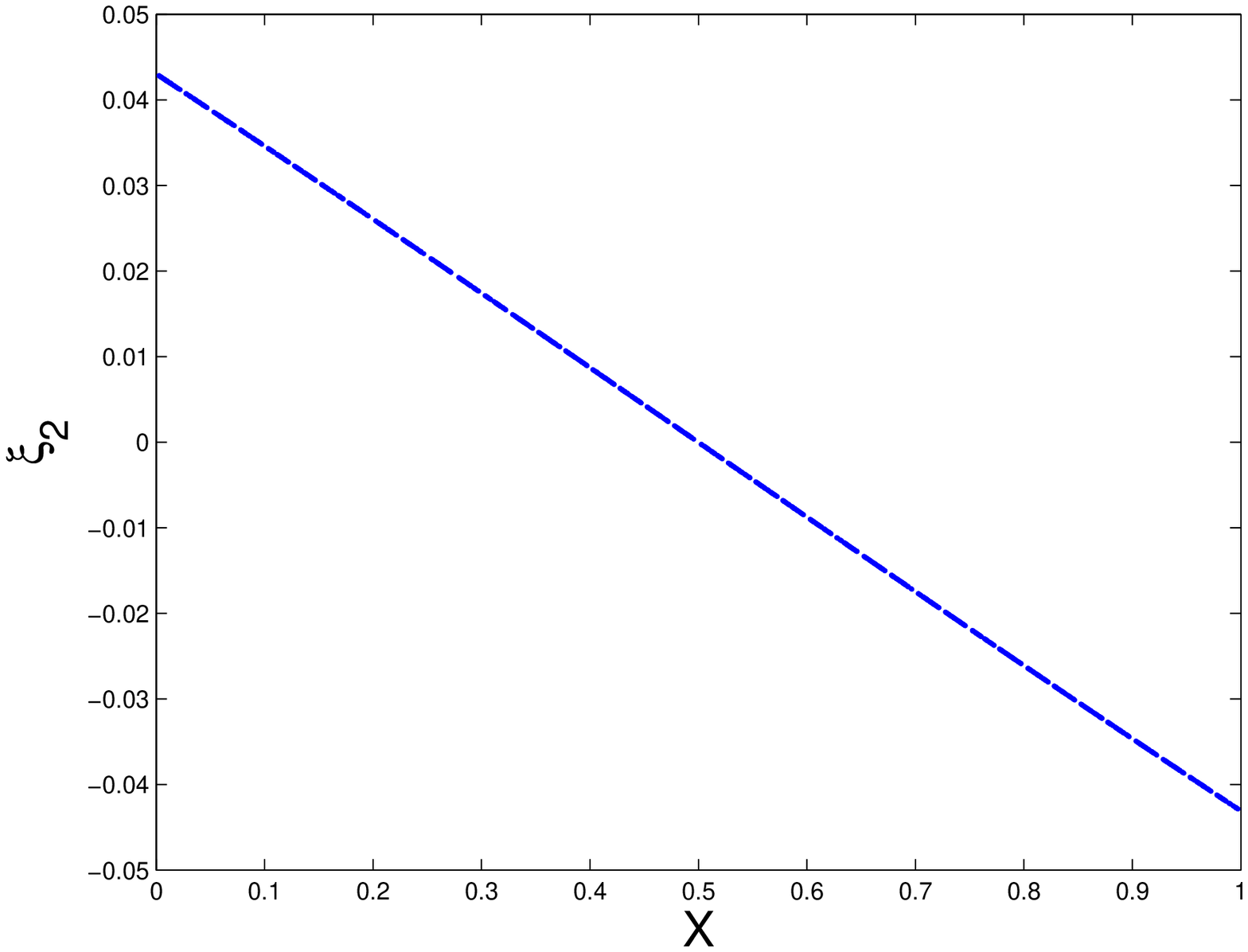}
   \includegraphics[width=.45\textwidth]{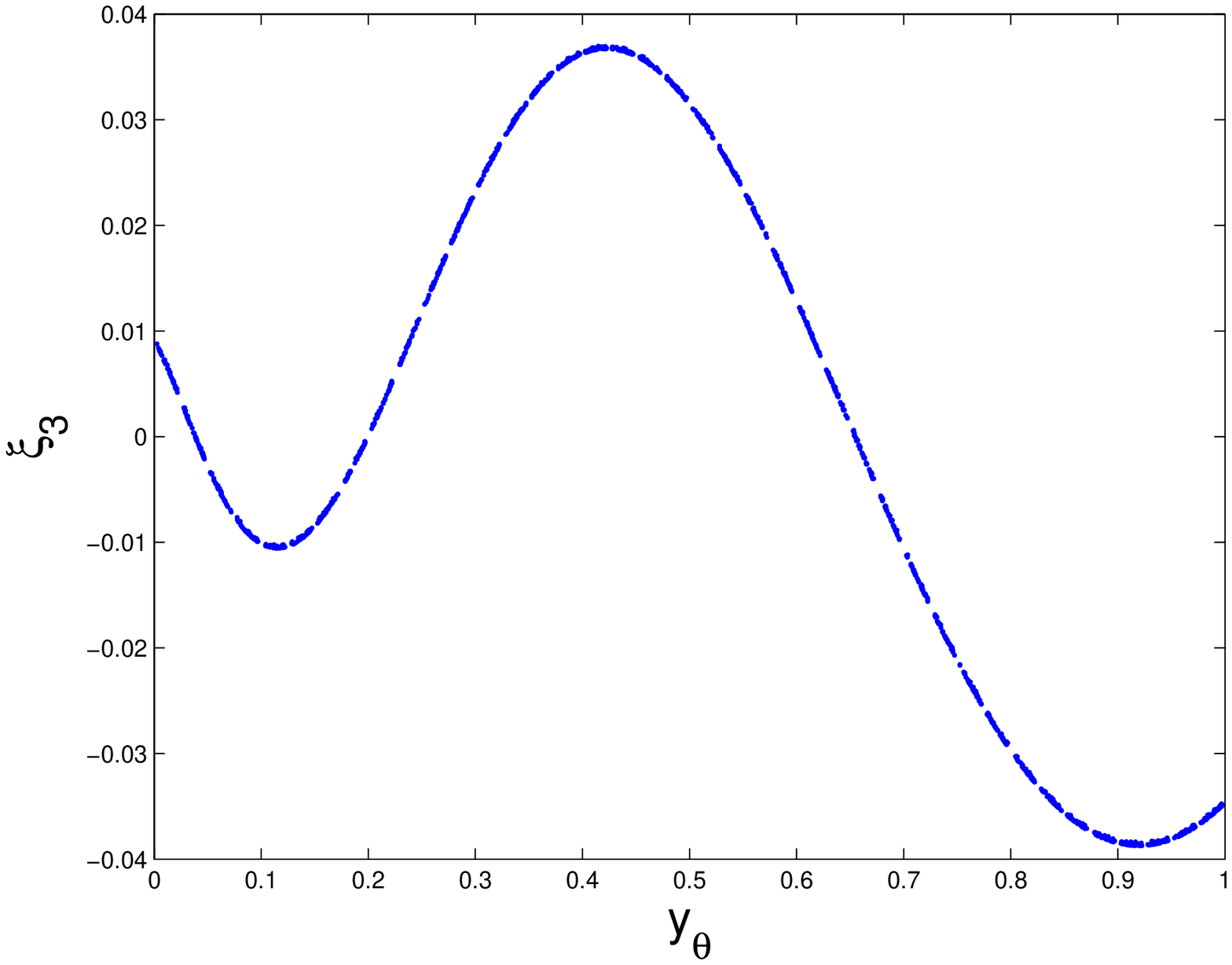}}
  \caption{Diffusion map applied to data in \f{F1}(b) with $\epsilon = 1$.  (a)  $\xi_2$ versus $X$. (b)  $\xi_3$ versus $y_\theta$.}
\label{F3}
\end{figure}

We would like to get the benefits of a large $\epsilon$ - the linearity in \f{F3}(a) without the drawbacks illustrated in \f{F3}(b).  For that reason, we study the large $\epsilon$ limit of diffusion maps in the next section.

\section{Diffusion Maps in the Large $\epsilon$ Limit}\label{S2}

We are really interested in non-linear manifolds, but for expository purposes we start by assuming that the manifold is linear.
Set the (arbitrary) origin to be at the point mean.  This implies that
\eqn{mean}
\sum_{i=1}^n x_{q,i} = 0
\enn
for all $q$.  Since the manifold is linear, it is now a linear subspace, D.

Define ${\bf x}_q$ to be the vector whose $i$-th component is $x_{q,i}$ and ${\bf e} = [1, 1, \cdots, 1]^T$ so \e{mean} can be written as
\eqn{zeroC}
{\bf e}^T{\bf x}_q = 0
 \enn
for all $q$.  (Bold face lower-case roman letters will always refer to $N$-dimensional column vectors while non-bold face roman letters will refer to $s$-dimensional column vectors.)

Suppose $\epsilon$ is large enough that $(\max({\bf H}_{ij})/\epsilon)^4$ can be ignored.  In that case,
\eq
{\bf W}_{ij} \approx 1 -  \left ( \frac{{\bf H}_{ij}}{\epsilon}\right )^2
\en
The row sums of this matrix are
\eq
\sum_{j=1}^N {\bf W}_{ij} \approx N - \frac{1}{\epsilon^2} \sum_{j=1}^N {\bf H}_{ij}^2
\en
so
\eqn{Kdef}
{\bf K}_{ij} \approx  \frac{1}{N}(1  - \frac{1}{\epsilon^2}[{\bf H}_{ij}^2 -\frac{1}{N}\sum_{k=1}^N {\bf H}_{ik}^2])
\enn
In other words,
\eq
{\bf K} \approx \frac{1}{N}({\bf E} - \frac{{\bf F}}{\epsilon^2})
\en
where ${\bf E}$ is the matrix of all ones and
\begin{eqnarray}
{\bf F}_{ij} & = & {\bf H}_{ij}^2 - \frac{1}{N}\sum_{k=1}^N {\bf H}_{ik}^2  \nonumber \\
 & = & \sum_{q=1}^s [(x_{q,i}-x_{q,j})^2 - \sum_{k=1}^N (x_{q,i}-x_{q,k})^2/N]  \nonumber \\
 & = & \sum_{q=1}^s[x_{q,j}^2 - 2x_{q,i}x_{q,j} - \sum_{k=1}^Nx_{q,k}^2/N].
\label{Fdef}
\end{eqnarray}

Writing $c_q$ for $\sum_{k=1}^Nx_{q,k}^2$ and defining vector ${\bf v}_q$ to have $i$-th entries $x_{q,i}^2$, \e{Fdef} can be written as
\eqn{finalF}
{\bf F} = \sum_{q=1}^s[\frac{{\bf e}({\bf v}_q - c_q{\bf e})^T}{N}- 2{\bf x}_q{\bf x}_q^T].
\enn

The points ${\rm X}_i$ lie in the linear subspace D.  Chose a coordinate system for S such that the first $d$ coordinates span D.  Hence, for $q > d$ ${\bf x}_q = {\bf v}_q = 0$ and $c_q = 0$ so we can rewrite \e{finalF} as
\eqn{finalLa}
{\bf F} = \sum_{q=1}^d[\frac{{\bf e}({\bf v}_q - c_q{\bf e})^T}{N}- 2{\bf x}_q{\bf x}_q^T].
\enn

Note that if a left or right eigenvector, ${\bf v}$ of ${\bf K}$ given by \e{Kdef} satisfies ${\bf e}^T{\bf v} = 0$ then it is also an eigenvector of ${\bf F}$and {\em vice versa}.

\noindent {\bf Observation}

If the points $\{X_i\}$ are regularly placed in the linear subspace - that is is they lie on the points of a regular rectangular grid or its extension to higher dimensions - and if the coordinate system is aligned with this grid, then ${\bf x}_q, q = 1, \cdots, d$ are eigenvectors of ${\bf F}$ and hence ${\bf K}$.

This follows by construction.  In this case the vector ${\bf x}_q$ takes the form
\eqn{eigen}
{\bf x}_q = e_1\otimes e_2\otimes \cdots \otimes e_{q-1}\otimes z_q \otimes e_{q+1} \otimes \cdots \otimes e_d
\enn
where $e_i$ is a column vector of ones of length equal to the number of grid points in the $i$-th dimension and $z_q$ is the set of coordinate points in the $q$-th dimension.  Note that $e_q^T z_q = 0$ for all $q$ so that the ${\bf x}_q$ are mutually orthogonal and all orthogonal to ${\bf e}$.  Hence
\begin{eqnarray}
{\bf F}{\bf x}_q & = & \sum_{p=1}^d[\frac{{\bf e}({\bf v}_p - c_p{\bf e})^T}{N}- 2{\bf x}_p{\bf x}_p^T]{\bf x}_q  \nonumber \\
 & = & \sum_{p=1}^d\frac{{\bf e}{\bf v}_p^T}{N}{\bf x}_q -2({\bf x}_q^T{\bf x}_q){\bf x}_q.
\label{Fxq}
\end{eqnarray}

If $p \ne q$ then ${\bf v}_p^T{\bf x}_q = 0$ directly from \e{eigen} so it only remains to show that for this data,
\eq
{\bf v}_q^T {\bf x}_q = \sum_{i=1}^N x_{q,i}^3 = 0
\en
which is true by virtue of the regular spacing and ${\bf e}^T{\bf x}_q = 0$.

If the points are ``reasonably" distributed, then this property is approximately true.  In \f{F4}(a) we show the points plotted via their entries, $\xi_{2,i}$ and $\xi_{3,i}$ of the eigenvectors of ${\bf K}$ evaluated with $\epsilon = 1$ for the planar data in \f{F1}(a).  Because the eigenvalue computation finds the dominant axis as second leading eigenvector (after the eigenvector ${\bf e}$) and that tends towards one of the diagonals when the data is random, the square has been rotated.  To demonstrate that these are close to linearly dependent on the original coordinates, we computed the affine translation of this representation that was closest to the original data in \f{F1}(a) (that is, the transformation that minimized the sum of the squares of the distances between associated points) and then plotted the components of the eigenvectors $\xi_2$ and $\xi_3$ against the original coordinates, as shown in \f{F4}(b).  ($\xi_3$ is offset by 0.1 for clarity.)  As can be seen, the dependence is almost linear.  The average distance error was 0.0023.

\begin{figure}[h]
   \parbox{\textwidth}{
   \includegraphics[width=.45\textwidth]{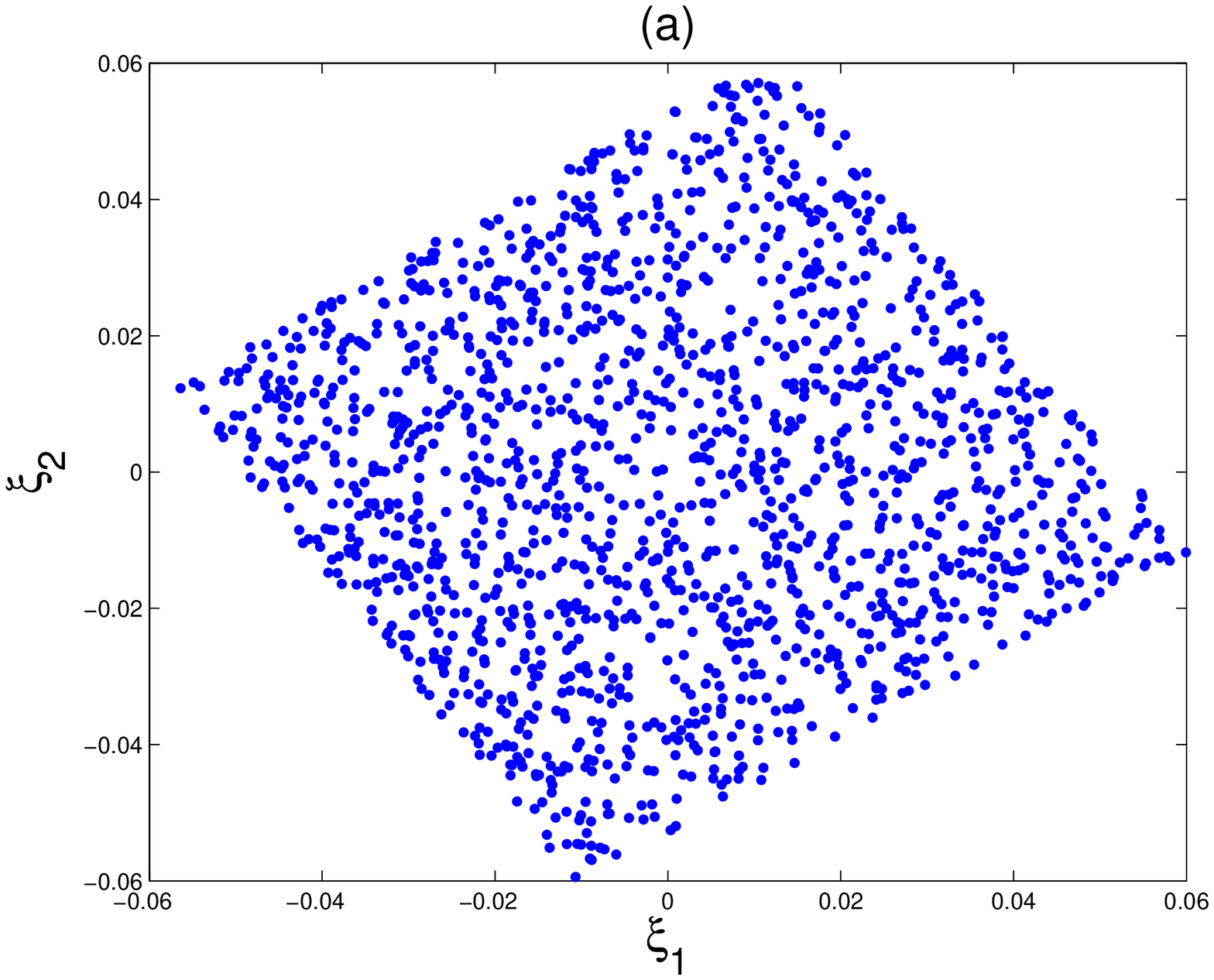}
   \includegraphics[width=.45\textwidth]{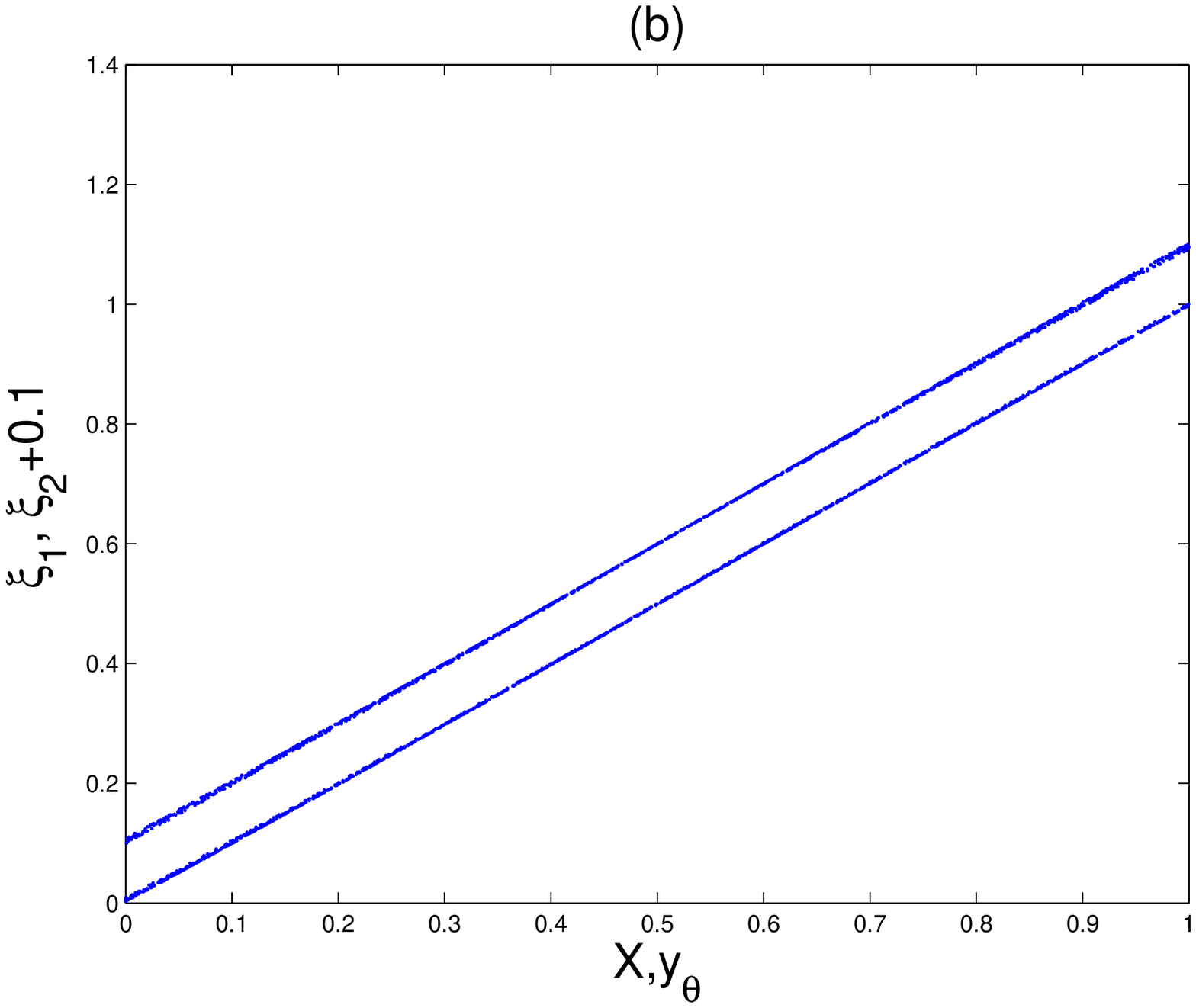} }
  \caption{(a) Points plotted by right eigenvectors of ${\bf K}$ with $\epsilon = 1$.  (b) Plot of eigenvector components in (a) after optimal affine transformation against coordinates in \f{F1}(a).}
\label{F4}
\end{figure}
However, it is more important to note that the left eigenvectors of ${\bf F}$ always have this property, regardless of the distribution of points, so we propose to use the left eigensystem of ${\bf F}$ but to avoid having to repeat ``left eigenvector" we will instead work with ${\bf G} = {\bf F}^T$ whose eigenvectors {\bf u} are the left eigenvectors of ${\bf F}$.

We are dealing with vectors in two different spaces: the $s$-dimensional original Euclidean space containing the points whose coordinates are ${\rm x}_i = [x_{1,i}, x_{2,i}, \cdots, x_{s,i}]^T$ and the $N$-dimensional space, P, containing the vectors ${\bf x}_q=  [x_{q,1}, x_{q,2}, \cdots, x_{q,N}]^T$.  If we pick any basis for the Euclidean space (which has to be orthogonal so that \e{Fdef} is based on the distance) it determines the entries in the $N$-dimensional vectors ${\bf x}_q$.  For each orthogonal set of vectors in the Euclidean space (which we will call {\em geometric} vectors) there is a corresponding set of vectors of the point coordinates (which we will call {\em point} vectors).
In particular, if we pick an (orthogonal) basis set for S such that the first $d$ basis vectors span D, then the $x_{q,i}$ satisfy $x_{q,i} = 0, q > d$.  Hence, the remaining $s-d$ basis vectors do not affect the $d$ vectors ${\bf x}_q, 1 \le q \le d$.  In this sense, to every orthogonal basis for D there is a corresponding set of $d$ linearly-independent point vectors in P.

To avoid verbal complexity, we will refer to ``non-zero eigenvectors" to mean ``eigenvectors corresponding to non-zero eigenvalues" and its obvious extensions.

We find that $G$ has the interesting properties summarized in\\
{\bf Theorem}

{\em If the components of ${\bf G}$ are given by}
\eqn{Gdef}
{\bf G} = \sum_{q=1}^d[\frac{({\bf v}_q - c_q{\bf e}){\bf e}^T}{N}- 2{\bf x}_q{\bf x}_q^T]
\enn
{\em then}
\begin{enumerate}
\item ${\bf G}$ has exactly $d$ non-zero eigenvalues.
\item The zero eigenvalue has algebraic multiplicity $N-d$ and geometric multiplicity at least $N-d-1$ although there are special cases where the geometric multiplicity is also $N-d$.
\item There exists a geometrically orthogonal basis for D such that the $d$ corresponding non-null point vectors ${\bf x}_q$ are the non-zero eigenvectors of ${\bf G}$.
\item These eigenvectors are orthogonal.
\item The corresponding eigenvalues are $-2{\bf x}_q^T{\bf x}_q$ so that all non-zero eigenvalues are negative.
\end{enumerate}

{\bf Proof}

We restrict ourselves to coordinate systems in S such that the first $d$ cordinates span M.  From \e{Gdef} we see that if {\bf b} is any vector orthogonal to {\bf e} and ${\bf x}_q,~q = 1, \cdots, d$ then ${\bf Gb} = 0$ so {\bf b} is a zero eigenvector.  Hence any $s-d-1$ (linearly independent) vectors that span the space orthogonal to {\bf e} and ${\bf x}_q,~q = 1, \cdots, d$ are independent zero eigenvectors and are orthogonal to {\bf e}.  Note that ${\bf e}^T$ is a zero left eigenvector of ${\bf K}$.  Since it is orthogonal to the $s-d-1$ zero right eigenvectors just enumerated, there must be one more zero eigenvalue.  (Usually its right eigenvector is a generalized eigenvector but there are special cases when this is not so, for example, when $N-1 = d = s$.)

We complete the proof by showing that there are $d$ independent non-zero eigenvectors.

Let ${\bf X}$ be the matrix whose columns are ${\bf x}_q, q = 1, \cdots, d$, and let ${\bf y} = {\bf X\alpha}$ be a linear combination of those vectors where ${\bf \alpha}$ is a $d$-dimensional vector.  Hence from \e{Gdef} and ${\bf e}^TX = 0$ we have
\eqn{trialeig}
{\bf G}{\bf y} = -2{\bf X} {\bf X}^T{\bf X}{\bf \alpha}
\enn
Define
\eqn{Ydef}
Y = {\bf X}{\bf X}^T.
\enn
It is a $d$ by $d$ symmetric matrix so has real eigenvalues and mutually orthogonal eigenvectors.  Because the points are in the $d$-dimensional subspace D and no lower dimensional subspace, $Y$ is positive definite so all of its eigenvalues are strictly positive.  Let ${\bf \alpha}$ be one of the eigenvectors with corresponding eigenvalue $\mu$.  Hence
$$
{\bf G}{\bf y} = -2\mu {\bf X} \alpha = -2\mu {\bf y}
$$
so ${\bf y} = {\bf X}\alpha$ is an eigenvector of {\bf G} with eigenvalue $-2\mu$.  This provides a complete set of $d$ non-zero eigenvectors.  If $Q$ is an orthonormal matrix whose columns are the (scaled) eigenvectors, $\{{\bf \alpha}\}$, of $Y$ then $Q$ can be applied to the initially selected Cartesian coordinate system for D to get a new Cartesian coordinate system with property 3.

Property 4 follows because the action of ${\bf G}$ defined by \e{Gdef} on the subspace spanned by $\{{\bf x}_q\}$ is exactly the same as the action of
\eqn{Ghat}
\hat{\bf G} = -2\sum_{q=1}^d{\bf x}_q{\bf x}_q^T
\enn
$\hat{\bf G}$ is a symmetric matrix so its eigenvectors are orthogonal.
Property 5 follows from \e{Gdef} when we choose a basis for D such that the eigenvectors are $\{{\bf x}_q\}$ (we have demonstrated how to do above).  Then $Y$ is diagonal with entries ${\bf x}_q^T{\bf x}_q$ which are its $d$ eigenvalues $\mu_q$.

As a footnote to this proof we observe that the number of non-zero eigenvalues follows quickly from a paper of nearly 75 years ago by Young and Householder\cite{YoHo38}.  They show that the matrix
\eq
\left [ \begin{array}{c c}
{\bf H} & {\bf e}\\
{\bf e}^T & 0 \end{array} \right ]
\en
has rank $d+2$ (see eq. (2) in their paper).  If we subtract the average of the first $N$ rows from the last and then add the new last row to each of the first $N$ rows (not changing the matrix rank) we get the matrix
\eq
\left [ \begin{array}{c c}
{\bf E}+{\bf G} & {\bf 0}\\
{\bf z}^T &-1 \end{array} \right ]
\en
where {\bf z} is some $N$-dimensional vector whose value does not affect the rank.  Hence, ${\bf E}+{\bf G}$ has rank $d+1$.  Since ${\bf e}^T$ is a left eigenvector of ${\bf G}$ and ${\bf E}$ with eigenvalues 0 and 1 respectively and ${\bf E}$ has rank 1,  the rank of ${\bf G}$ is $d$.

\section{Modifying Distances to get a Good Parameterization}\label{S3}

We see from \e{Gdef} that the non-zero eigenvectors of ${\bf G}$ are the eigenvectors of $XX^T$ and these are precisely the principal components from PCA so the method proposed in \s{S2} provides exactly the PCA representation of the data set, which raises the question ``What is the advantage of this method over PCA?"  PCA starts with the coordinates of the data whereas this method starts with the distances of each point pair (and incidentally provides the coordinates in an orthogonal coordinate system from the calculation).  A potential application of this method is to a set of objects that are not given in a Euclidean space but for which a similarity distance can be defined (e.g., a set of graphs).  In this report, we focus only on sets that can be represented as points in a Euclidean space.

The problems with the diffusion map parameterization (using large $\epsilon$ as discussed in \s{Sdemo}) arose when distances that were large in the manifold were small in the original space.  Because the algorithm only uses distances we can consider modifying the distances of more distant objects to be approximately compatible with those in the manifold.

A procedure for doing this is as follows:
\begin{enumerate}
\item Remove all edges from the graph representation of the distance matrix ${\bf H}$ longer than some value.
\item Compute the new shortest path to all edges to get a modified distance matrix
\item Use this modified matrix to generate a parameterization using the technique of \s{S2}
\end{enumerate}

In \f{F5}(a) we show the points plotted in the $(\xi_1, \xi_2)$-plane using the modified distance matrix approach for the Swiss roll data in \f{F1}(b).  Edges were removed from the graph that were longer than $2\beta$ where $\beta$ was the minimum value that did not lead to a disconnected graph when all larger edges were cut ($\beta$ is 0.0322).  As happened in \f{F4}, the square is twisted.  As before, we found the best affine transformation of these points to match the data in \f{F1}(a). The new coordinate systems is called $\phi_1$ and $\phi_2$. In \f{F5} (b) and (c) we show the plots of $X$ versus $\phi_1$ and $y_\theta$ versus $\phi_2$.  As can be seen, the relationship is almost linear.

\begin{figure}[h]
   \parbox{\textwidth}{
   \hspace{.22\textwidth}
   \includegraphics[width=.45\textwidth]{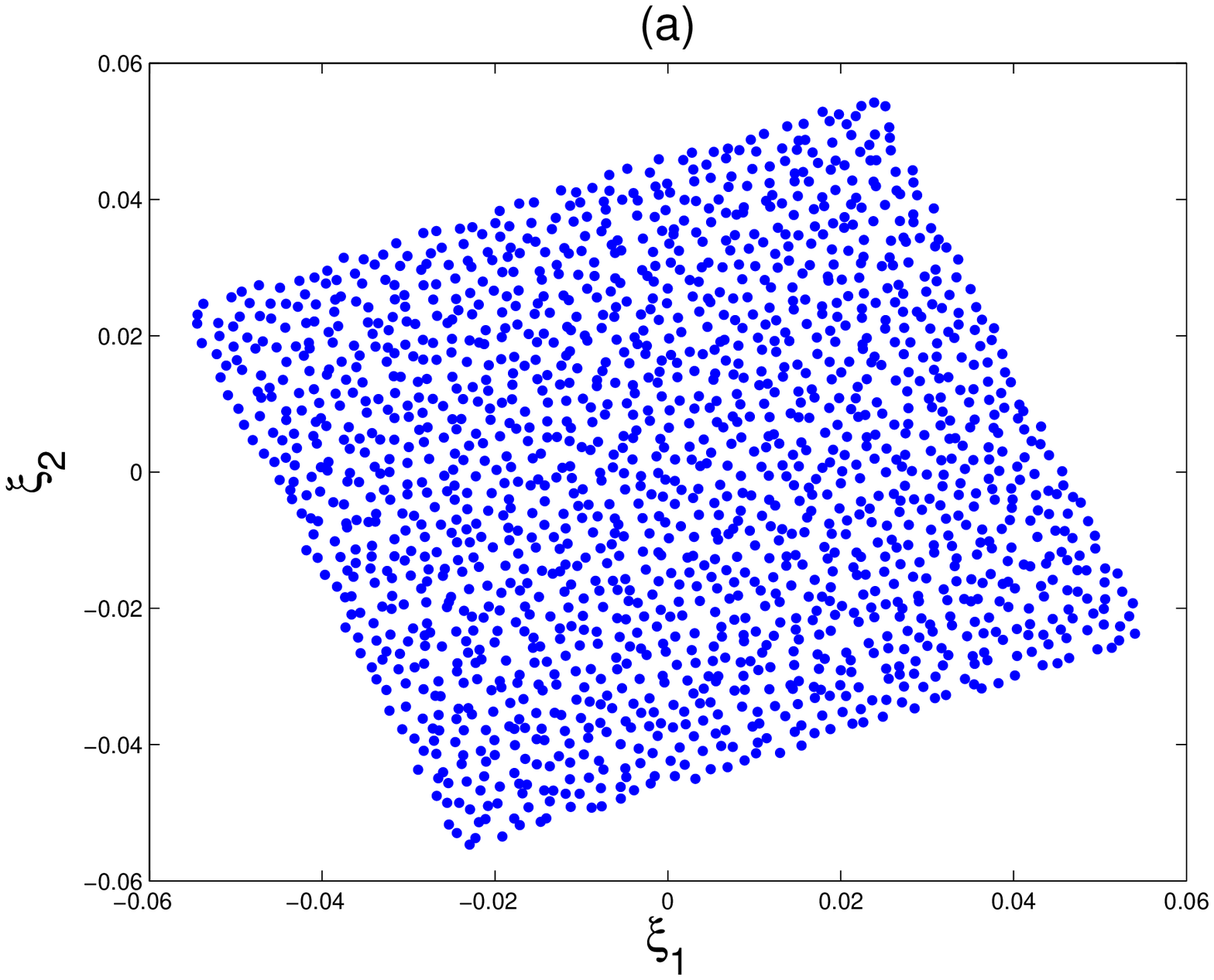}}
   \parbox{\textwidth}{
   \includegraphics[width=.45\textwidth]{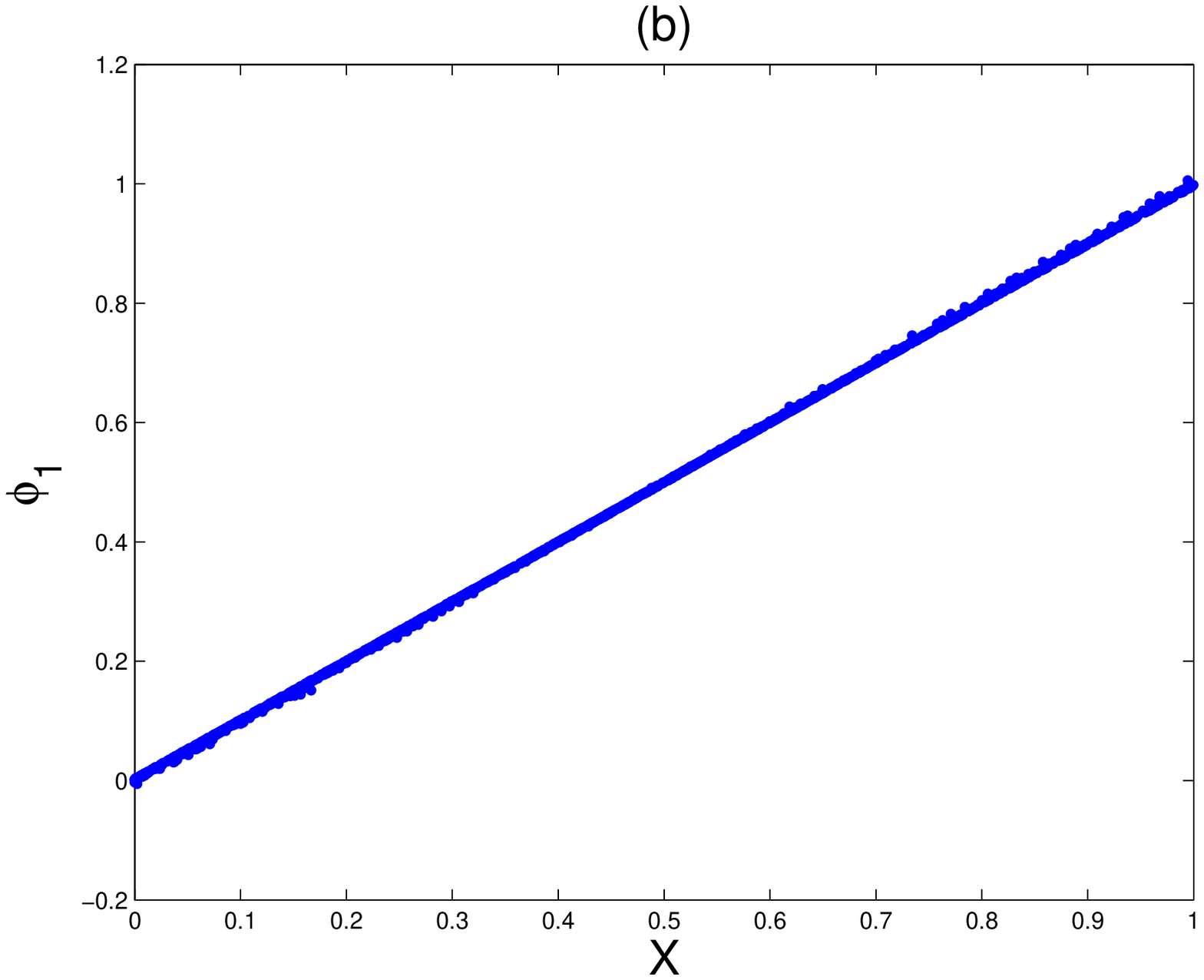}
   \includegraphics[width=.45\textwidth]{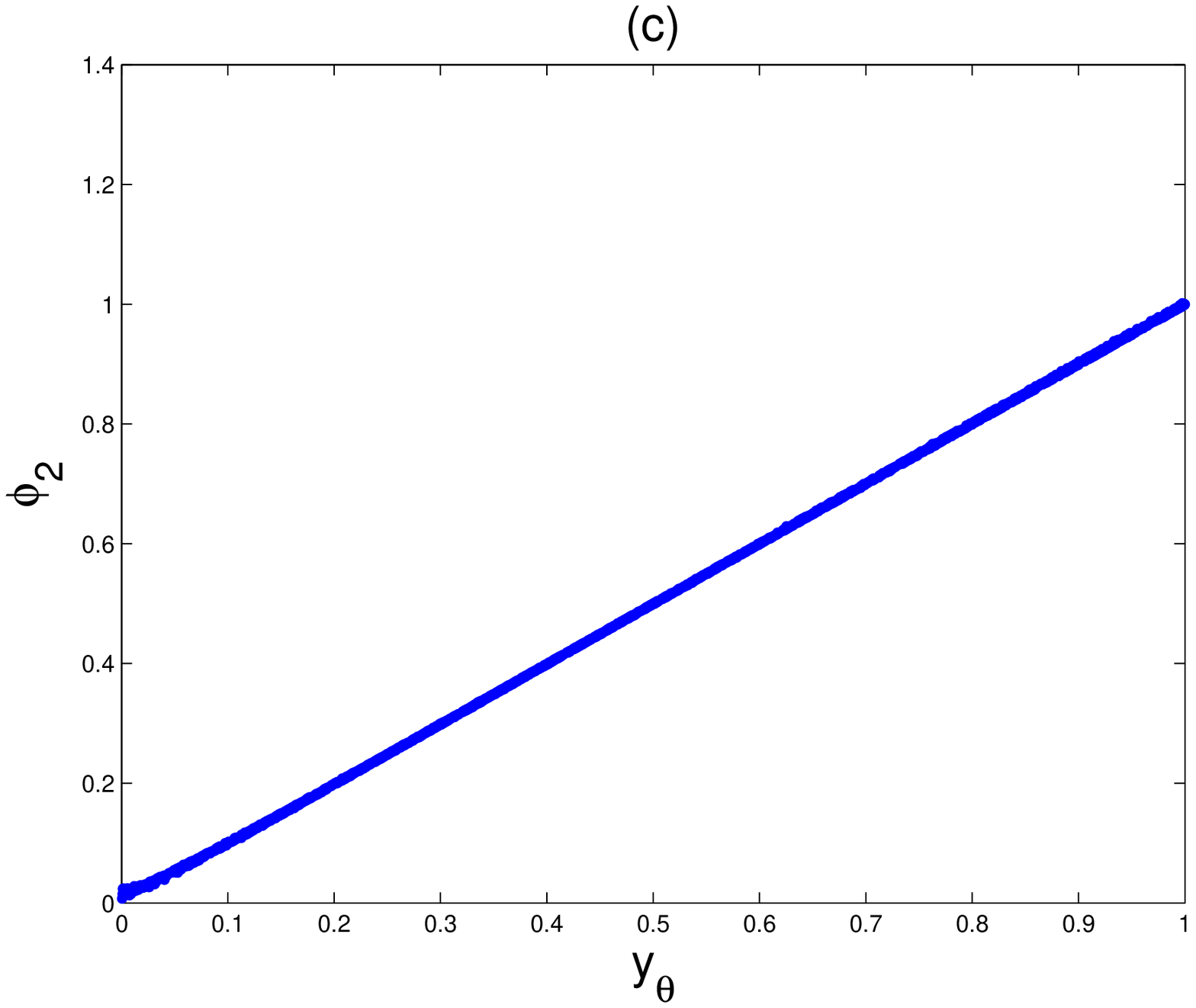}}
  \caption{Modified distance methods applied to Swiss roll data. (a) Points plotted by right eigenvectors of ${\bf G}$.  (b) and (c) Plot of rotated coordinates of (a) against coordinates in original plane data.}
\label{F5}
\end{figure}

The distance modification scheme gives good results for the Swiss roll data because it was obtained by twisting a linear manifold without stretching.  However, if it is necessary to stretch the data to ``unfold" it into a linear manifold, the scheme will not work unless some of the nearest neighbor distances are also modified.  This can be seen with the example in \f{F6}(a) which consists of a set of points somewhat uniformly placed on the majority of a surface of a sphere.  In this example, the surface of the sphere is included from the ``North pole" to to Southern latitude $67.5^o$.  A point was placed at the North pole and then points were place on each of 15 circles of constant latitude at constant separations of $10.5^o$.  The points were equi-spaced on each circle.  40 points were placed on the equator and the number on each of the other circles was chosen to make the spacing on each circle as similar as possible except that a minimum of 6 points were placed on each circle.  The location of one of the points was chosen from a uniform random distribution.  In \f{F6}(a) all points are plotted as dots except for the points on the southern-most latitude which are circles.  Clearly this 2D surface can be parameterized easily.  The southern-most latitude circle is the boundary of the finite, non-linear manifold.

The results of the standard diffusion map and the modified distance scheme are shown in \f{F6} (b) and (c), respectively.  The plots shown the circles of constant latitude in the eigenvalue coordinates.  The southern-most circle is shown as a thicker line.

\begin{figure}[h]
   \parbox{\textwidth}{
   \hspace{.22\textwidth}
   \includegraphics[width=.45\textwidth]{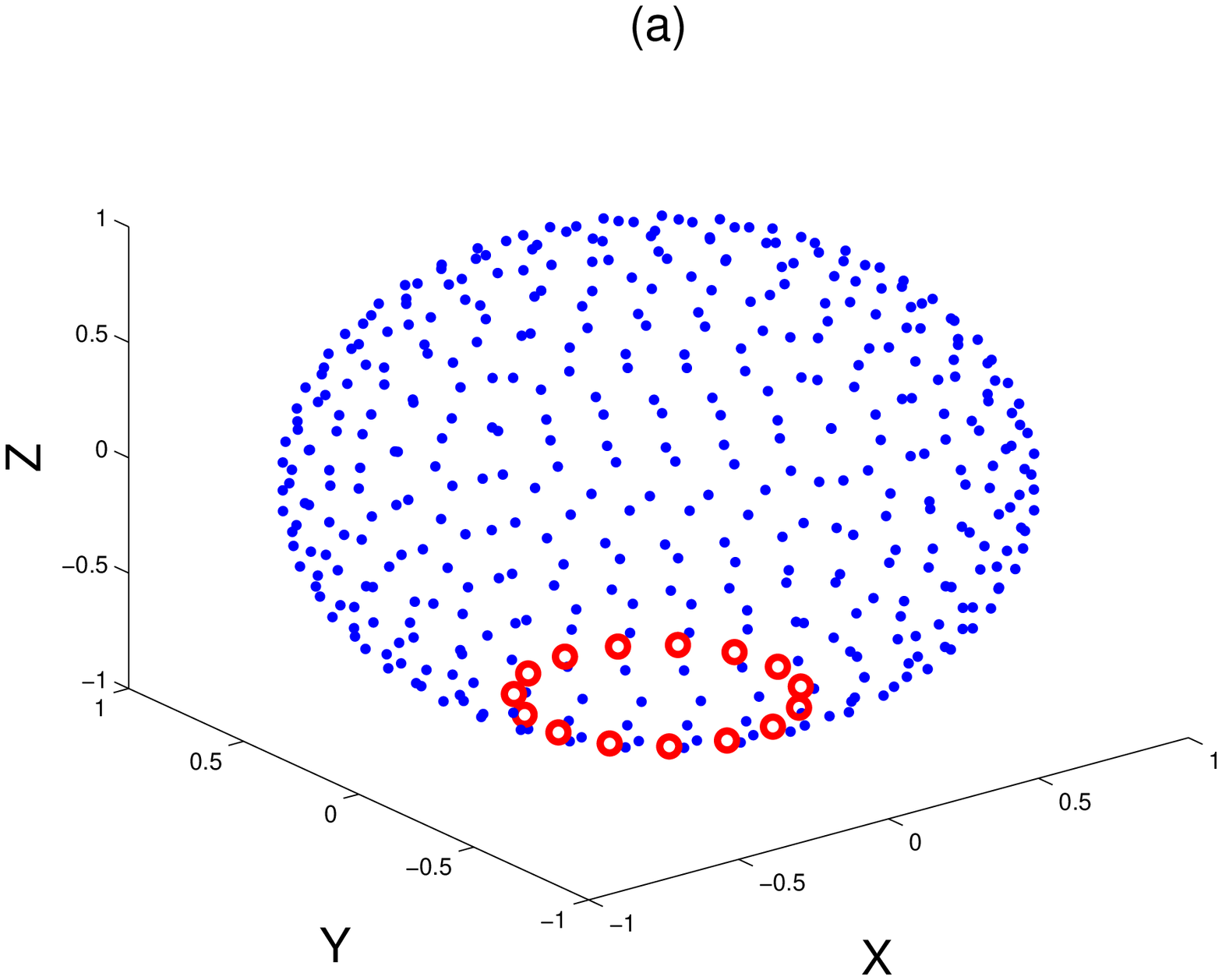}}
   \parbox{\textwidth}{
   \includegraphics[width=.45\textwidth]{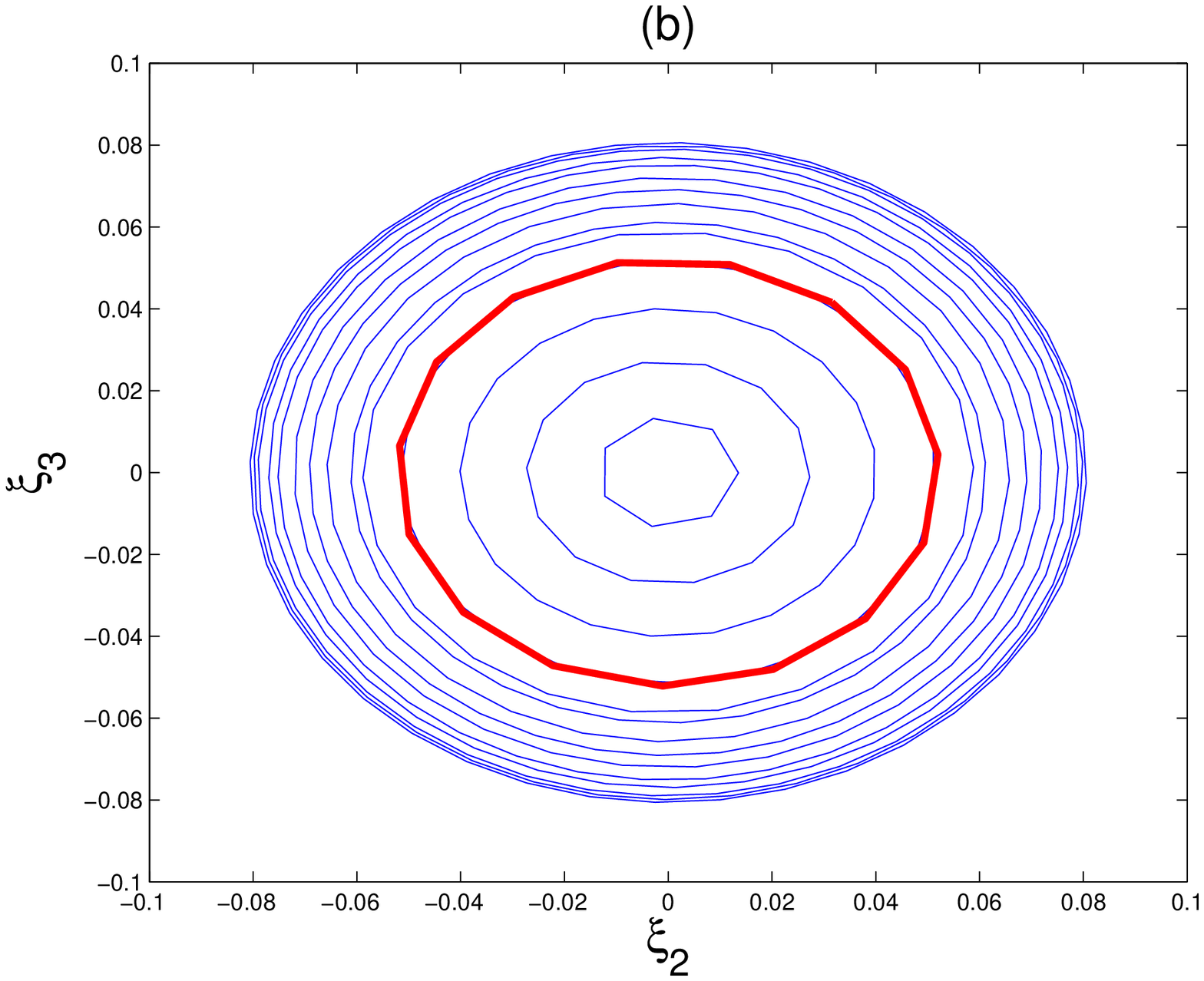}
   \includegraphics[width=.45\textwidth]{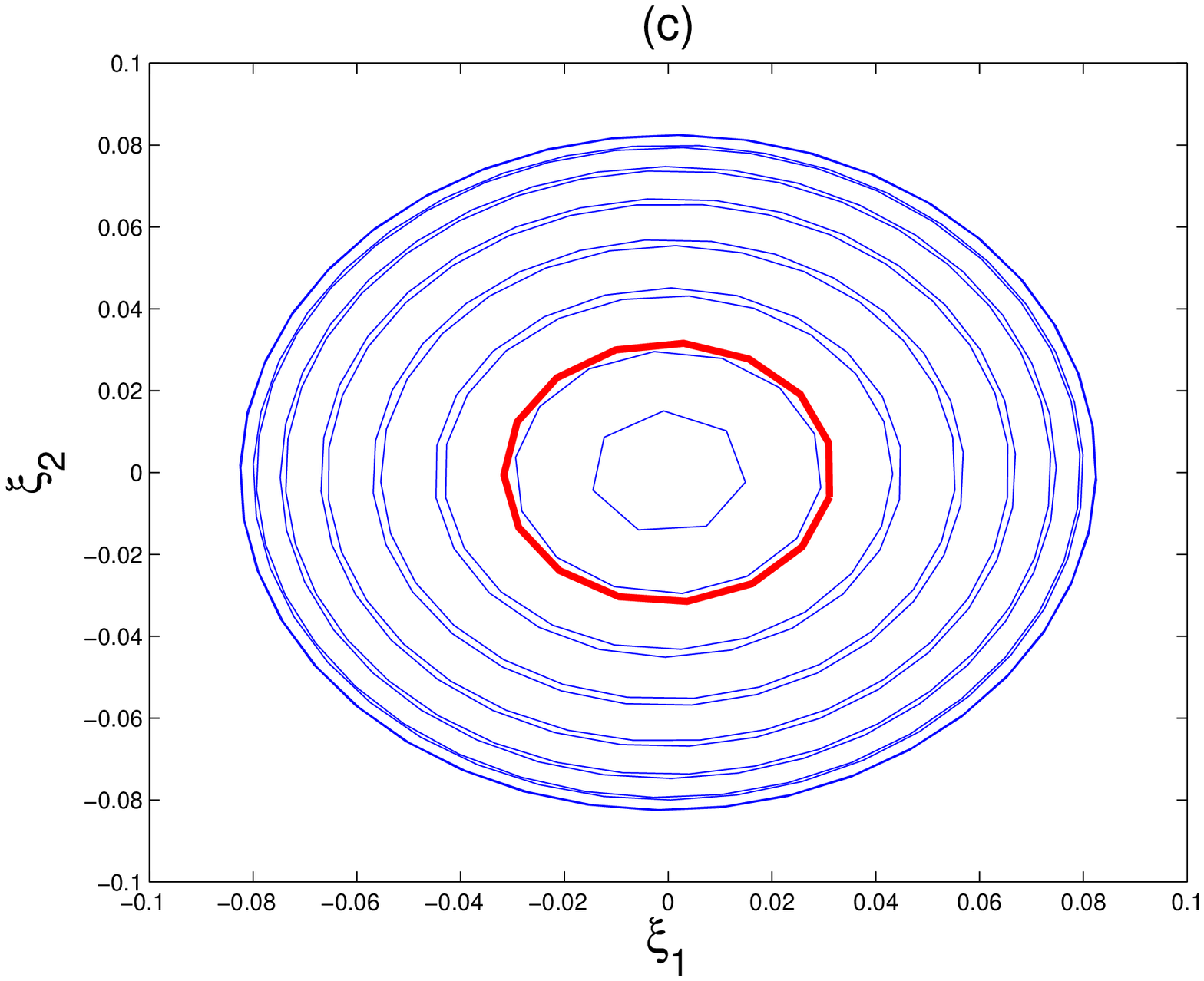}}
  \caption{(a) Original data on most of the surface of a sphere.  (b) Diffusion Map applied to this data.  (c) Modified distance scheme applied to same data.}
\label{F6}
\end{figure}

As can be seen from the figure, the modified distance scheme has the boundary of the manifold in the interior of parameterization it produces (as does the diffusion map process, although not as badly).  This arise because the scheme places a great emphasis preserving the length of nearby neighbors, and in this case there is a circle of nearby neighbors with a perimeter that is much smaller than the perimeter of other circles that are in the interior of the 2D manifold.  Clearly it is necessary to modify the lengths of some nearby neighbors to get a good parameterization.  A clue to the need for this can be obtained from the size of additional eigenvalues of ${\bf G}$ and local changes in the components of their eigenvectors.  Since the method is equivalent to a PCA decomposition\footnote{This is true if the distance matrix corresponds to a feasible configuration in real space.  If it doesn't, for example because the distance inequality is not satisfied, we will get some complex eigenvalues and eigenvectors}, other eigenvalues must be large when we find that points significantly separated in the original space are close in the parameter space.  How to systematically modify nearby neighbors distances in this case is a subject of ongoing research.




\begin{thebibliography}{10}

\bibitem{CoLa06}
R.~R. Coifman and S.~Lafon, \emph{Appl. Comput. Harmon. Anal.} \textbf{21}, pp. 5--30 (2006)



\bibitem{RoSa00}
S. T. Roweis and L. K. Saul, Nonlinear Dimensionality Reduction by Locally Linear Embedding, {\em Science}, {\bf 290}, pp 2323-2326, (2000)

\bibitem{Samm69}
J. W. Sammon Jr.,
A Nonlinear Mapping for Data Structure Analysis, {\em IEEE Trans. Computers}, {\bf C-18}, 5, pp 401-409, (1969)

\bibitem{SGSK12}
B. Sonday, C. W. Gear, A. Singer and I. G. Kevrekidis, Solving Differential Equations by Model Reduction on Learned Manifolds, see www.princeton.edu/$\thicksim$wgear/SGSK.pdf

\bibitem{SuCrFy11}
J. Sun, M. Crowe, and C. Fyfe,
Extending Metric Multidimensional Scaling with Bregman Divergences, {\em Pattern Recognition}, {\bf 44}, pp 1137-1154, (2011)

\bibitem{SuFyCr12}
J. Sun, C. Fyfe, and M. Crowe,
Extending Sammon Mapping with Bregman Divergences, {\em Information Sciences}, {\bf 187}, pp 72-92, (2012)

\bibitem{YoHo38}
G. Young and A. S. Householder,
Discussion of a Set of Points in Terms of their Mutual Distances, {\em Psychometrica}, {\bf 3}, 1, pp 19-21, (1938)

\bibitem{ZhZh05}
Z. Zhang and H. Zha, Principal Manifolds and Nonlinear Dimension Reduction via Local Tangent Space Alignment, {\em SIAM Journ. on Scientific Computing}, {\bf 268}, 1, pp 313-338, (2005)

\end{thebibliography}
\end{document}